\documentclass[aps,preprint,tightenlines,epsfig,preprintnumbers]{revtex4}
\newcommand{\eq}[1]{Eq.~(\ref{#1})}
\newcommand{\be}{\begin{equation}}
\newcommand{\ee}{\end{equation}}
\newcommand{\bea}{\begin{eqnarray}}
\newcommand{\eea}{\end{eqnarray}}
\newcommand\mn{{\mu\nu}}

\begin{document}

\title{Role of the nuclear vector potential  in deep inelastic
  scattering}

\author{W. Detmold} \affiliation{Department of Physics,
  University of Washington, Seattle, WA 98195-1560, U.S.A.}
\author{G. A. Miller} \affiliation{Department of Physics,
  University of Washington, Seattle, WA 98195-1560, U.S.A.}
\author{J. R. Smith} \affiliation{Department of Physics,
  University of Washington, Seattle, WA 98195-1560, U.S.A.}

\begin{abstract}
  We study the influence of the strong nuclear vector potential,
  treated using the mean-field approximation, in deep inelastic
  scattering.  A consistent treatment of the electromagnetic current
  operator, combined with the use of the operator product expansion is
  presented and discussed.
\end{abstract}

\preprint{NT@UW-05-10}
\pacs{25.30.Mr, 11.80.-m,12.39.Fe,13.60.-r}

\maketitle


  





  


\section{Introduction}
The EMC effect revealed that the structure of nucleons bound within
the nucleus differs from those of free space \cite{EMCE,EMCT}.
Indeed, careful treatments \cite{JAF}-
\cite{MIL} of Fermi motion and binding effects of nucleons cannot
explain the observed reduction of the nuclear structure function, or
distribution function $q(x)$, in the range of Bjorken $x$ between 0.3
and 0.8.  Therefore it is worthwhile  to derive models of nuclei in
which the internal quark structure of nucleons responds to the nuclear
environment. One may then investigate whether such responses account
for the observations.

This is a daunting task. One way to proceed
\cite{Thomas:vt}-\cite{millersmith} is to use a mean field models in
which quarks in a given nucleon feel the influence of mesons produced
by the average field of the nucleus.  Such models should account for
the saturation properties of nuclear matter, as well as the structure
functions of the free nucleon.  Early work on this
problem~\cite{Thomas:vt} was based on the quark-meson coupling model
\cite{Guichon:1987jp} in which quarks interact by the exchange of
scalar and vector mesons.  More recently, a quark-diquark description
of the single nucleon~\cite{MIN1,MIN2}, based on the NJL
model~\cite{NJL}, was combined with the mean field description of
nuclear matter \cite{BT,QM}.  Another set of work is based on a
nuclear matter version~\cite{millersmith} of the chiral quark soliton
model of Diakonov {\it et al.}~\cite{cism}.  Here, the nuclear attraction is
generated by the exchange of pairs of pions (yielding an effective
scalar potential) with the environment, and the repulsion arises from
vector meson exchange.

A general feature of these mean-field models is that the binding
interactions can be expressed in terms of scalar and vector
potentials. The treatment of the scalar potentials is straightforward,
but the vector potential is more subtle.  In present mean-field models
the vector potential is a constant that changes the energy of a
nucleon, but causes no change in the wave function.  It has been
argued that the quark struck by the hard photon should not feel the
nucleon vector potential $3V^0$, and accounting for this causes a
shift in the argument of the distribution function, and a change in
its normalization.  This shift seems to have a significant effect on
computed numerical results \cite{Thomas:vt,Mineo:2003vc}, but its
presence is not immediately apparent in other work \cite{millersmith}.

The sole aim of the present note is to clear up the technical issue of
whether it is necessary, if one is using a mean field approximation,
to include this shift in the argument To focus on the main point we
simplify and use infinite nuclear matter (in which the scalar and
vector potentials are treated as constants), examine only the
spin-averaged structure function, and ignore QCD radiative effects by
assuming the { Bjorken\/} limit.

We proceed by discussing the standard derivation of the parton model
\cite{JAF,Jaffe95} for a free nucleon, and then immerse the nucleon in
the medium. The paper ends by providing an interpretation that unifies
the approaches \cite{Thomas:vt,Mineo:2003vc} and \cite{millersmith}.

\section{Free Nucleon}

Consider charged lepton scattering on a free nucleon in which the
initial lepton exchanges a photon of momentum $q$ with a target of
momentum $P$.  The differential cross-section for inclusive scattering
depends on the hadronic tensor $W^{\mu\nu}$:
\begin{eqnarray}
        4\pi W^{\mu \nu} & = &
           \int d^4\xi e^{iq\cdot \xi}
        \langle PS\vert [ J^\mu(\xi) ,
        J^{\nu}(0) ] \vert PS \rangle _c, 
        \label{eq:com} 
\end{eqnarray}
expressed in terms of connected ($_c$) matrix elements of the
electromagnetic current operator $J^\mu$:
\bea
        J_{\mu}(\xi) = \bar{\psi}(\xi) \gamma_\mu \hat{Q}
        \psi(\xi).\label{current}
\eea
The charge operator is $\hat{Q}$ and the states are covariantly
normalized to: $ \langle P\vert P^\prime \rangle = 2E (2\pi )^3
\delta^3 (P-P^\prime).  $ Using Lorentz covariance, gauge invariance,
parity conservation in electromagnetism and standard discrete
symmetries of the strong interactions, $W^{\mu \nu}$ can be
parametrized in terms of four scalar dimensionless structure functions
$F_1(x,Q^{2})$, $F_2(x,Q^{2})$, $g_1(x,Q^{2})$ and $g_2(x,Q^{2})$. We
shall be concerned with spin averaged quantities, and keep only the
symmetric part of $W^\mn$:
\begin{eqnarray}
        W_s^{\mu \nu} 
& \equiv 
        & \left( -g^{\mu \nu}+{q^{\mu}q^{\nu}\over q^2}\right) 
        F_1 +\left[ \left( P^{\mu}-{\nu \over q^2}q^{\mu} \right) 
        \left( P^{\nu}-{\nu \over q^2}q^{\nu} \right) \right] {F_2 \over \nu}, 
        \label{eq:WS}
\end{eqnarray}
We use Jaffe's conventions~\cite{Jaffe95} in which $W^{\mu\nu}$ is
dimensionless.

We evaluate $F_1(x)=W_s^{11}$ in the parton model for a free nucleon
using the laboratory system in which
$q^\mu=\{q_0,0_\perp,-\sqrt{q_0^2+Q^2}\}$ and $P^\mu=\{M,0_\perp,0\}.$
Using light-cone coordinates, $q\cdot\xi=q^+\xi^-+q^-\xi^+$, with
$q^+=-Mx/\sqrt{2}$ and $q^-=2q^0+{Q^2\over2q^0}$ in the Bjorken limit.

The leading contribution in the operator product expansion of the
cross section (or forward Compton amplitude) is the handbag diagram,
obtained by evaluating the current commutator and keeping the most
singular terms \cite{Jaffe95}.  The identity
\bea
        [\bar\psi_1\psi_1,\bar\psi_2\psi_2]
        =\bar\psi_1\{\psi_1,\bar\psi_2\}\psi_2-
        \bar\psi_2\{\psi_2,\bar\psi_1\}\psi_1
\label{commie}\eea
is useful. This is obtained by neglecting the (unequal-time)
anti-commutator of the $\psi$ fields, and is valid for non-interacting
quarks or for quarks immersed in a constant background field.  For a
massless field,
\bea
        \{\psi(\xi),\bar\psi(0)\}=
        {1\over 2 \pi} \partial \hspace{-2mm}/ \epsilon (\xi_0) \delta (\xi^2),
\label{green}
\eea
(with $\epsilon(\xi_0)=1 $ for $\xi_0\ge0$ and $\epsilon(\xi_0)=-1 $
for $\xi_0<0)$.  The relevant current commutator is then given by 
\bea && \left[ J^{1 } (\xi), J^{1 } (0) \right]= -{1\over 2 \pi}[
\bar\psi(\xi)\widehat{Q}^2
\left(
  (\partial^1\gamma^1+\partial^1\gamma^1-g^{11}\gamma\cdot\partial)
  \epsilon (\xi_0) \delta (\xi^2) \right) \psi(0)\nonumber 
\\&& -
\bar\psi(0)\widehat{Q}^2 \left(
  (\partial^1\gamma^1+\partial^1\gamma^1-g^{11}\gamma\cdot\partial)
  \epsilon (\xi_0) \delta (\xi^2) \right) \psi(\xi)].
        \label{eq:puttel}
\eea
The terms $\partial^1\gamma^1$ are of the size of small momenta and
can be ignored in the Bjorken limit \cite{JAF}.  The surviving term:
$\gamma\cdot\partial$ can therefore be replaced by
$\gamma^+\partial^-+\gamma^-\partial^+$.  Then one may evaluate the
integral over $d^4\xi$.  The term $\gamma^-\partial^+$ is
exponentially suppressed by the large momentum $q^-$ and is ignorable.
The result is the standard parton model:
\bea 
F_1(x)= {1\over
  2\sqrt{2}\pi}\int d\xi^-e^{-iMx\xi^-\over\sqrt{2}}
\langle P\mid\psi^\dagger(\xi^-) \hat{Q}^2P_+
\psi(0)-
\psi^\dagger(0) \hat{Q}^2P_+\psi(\xi^-)\vert P\rangle
\LARGE{\vert}_{\xi_\perp=0,\xi^+=0},\;
\label{reduce}
\eea
with $P_+= {1\over2}(1+\alpha^3)$.  Thus the parton model result
emerges by taking the current commutator and keeping leading
singularities \cite{leading} along the light cone \cite{Jaffe95}.  In
model calculations, one evaluates \eq{reduce} using wave functions
determined at a low momentum scale, $Q_0^2$. Then QCD evolution is
used to obtain distributions to compare with deep inelastic scattering
data.

\newcommand{\bold}[1]{\mbox{\boldmath ${#1}$}}
\newcommand{\bfxi}{\bold{\xi}}
\section{Bound nucleon}

The next step is to immerse the nucleon in the nucleus.  There are
some general features that are common to both sets of models discussed
in the introduction.  In each model quarks satisfy a mode equation of
the general form
\bea \left(-i\bold{\gamma}\cdot\bold{\nabla} +m(r)\right) q_n(\bfxi)=(E_n-V^0)q_n(\bfxi)
,\eea
with the position dependent constituent quark mass, $m(r)$ accounting
for the influence of internal binding potentials and interactions with
scalar objects produced by the surrounding medium.  This can be
obtained by solving an in-medium gap equation for $m$ or a
self-consistency condition. The quark is also influenced by a constant
vector potential $V^\mu$, with the time component the only non-zero
component for nuclear matter at rest.  The time dependence of each
mode is given by $e^{-iE_n\xi^0}$ with the vector potential
contributing a factor $e^{-iV^0\xi^0}=e^{-i(V^+\xi^-+V^-\xi^+)}$.  The
field operators, now denoted as $\Psi$, can be expressed in terms of
these mode functions that embody the influence of the medium.  We now
repeat the derivation of Eqs.~(\ref{eq:com})-(\ref{reduce}), using
these new field operators. The steps from \eq{eq:com}-\eq{commie} are
as before, but the result (\ref{green}) is changed to 
\bea
\{\Psi(\xi),\bar\Psi(0)\}=e^{-iV\cdot\xi} {1\over 2 \pi} \partial
\hspace{-2mm}/ \epsilon (\xi_0) \delta (\xi^2).
\label{green1}
\eea
Note the appearance of new phase factor.  Using this and a similar
expression for $\{\Psi(0),\bar\Psi(\xi)\}$ one finds an in-medium
version of the structure function, $\widetilde{F}_1$: 
\bea
\label{reduce1}
 \widetilde
{F}_1(x)&=& {1\over 2\sqrt{2}\pi}\int d\xi^-e^{-iP^+\xi^-}
\langle P\mid e^{-iV^+\xi^-}\Psi^\dagger(\xi^-) \hat{Q}^2P_+
\Psi(0)
\\
&&\hspace*{5cm}
- e^{iV^+\xi^-}\Psi^\dagger(0) \hat{Q}^2P_+\Psi(\xi^-)\vert P\rangle
\LARGE{\vert}_{\xi_\perp=0,\xi^+=0},
\nonumber
\eea 
with $P^\mu$ the nucleon momentum.  At first glance, the presence of
the phase factors $e^{\mp iV^+\xi^-}$ seems to cause this result to
differ substantially from that of \eq{reduce}. However the constant
vector potential discussed above causes the mode functions to have a
phase that cancels these phase factors.  Thus the expression for the
in-medium structure function contains no additional phase factors.

This can be expressed more formally using the techniques of
Ref.~\cite{Mineo:2003vc} who observe that the quark Hamiltonian in the
mean field approximation for nuclear matter at rest has the form
\begin{eqnarray}
\hat{H}_q = \hat{h}_q + V_0\,\hat{N}, \label{hq}
\end{eqnarray}
where $\hat{h}_q$ is the quark Hamiltonian in the absence of the mean
vector field, and $\hat{N} = \int {\rm d}^3x\,\Psi^{\dagger}(x)
\Psi(x)$ is the quark number operator.  Ref.~\cite{Mineo:2003vc} then
uses translation invariance to obtain the relation,
\begin{equation}
\Psi(\xi) = e^{i {\hat P}_q\cdot \xi} \Psi(0) e^{-i {\hat P}_q\cdot \xi}\,,  \label{heis}
\end{equation}
where ${\hat P}_q^{\mu}=({\hat H}_q, {\hat {\bold P}}_q)$ the
4-momentum operator for quarks.  This leads to:
\begin{equation}
\Psi(\xi) = e^{-i V \cdot \xi} \Psi_{0}(\xi).  \label{gauge}
\end{equation}
Here $\Psi_{0}$ is the quark field in the absence of the vector
potential (but influenced by the medium via scalar interactions).
Using \eq{gauge} in \eq{reduce1} allows us to obtain 
\bea \widetilde
{F}_1(x)= {1\over 2\sqrt{2}\pi}\int d\xi^-e^{-iP^+\xi^-}
\langle P\mid \Psi_0^\dagger(\xi^-) \hat{Q}^2P_+
\Psi_0(0) - \Psi_0^\dagger(0) \hat{Q}^2P_+\Psi_0(\xi^-)\vert P\rangle
\LARGE{\vert}_{\xi_\perp=0,\xi^+=0}.\;
\label{reduce2}
\eea 

Equation (\ref{reduce2}) tells us that the vector potential causes no
shift in the argument.  This follows from using the fields $\Psi$ {\it
  everywhere} in the electromagnetic current and consistently
evaluating the current commutator.  Actually, the result
(\ref{reduce2}) may be obtained immediately by noting that the current
operator (\ref{current}) expressed in terms of interacting fields
$\Psi$ can be re-expressed using (\ref{gauge}) in terms of the quark
field, $\Psi_0$, obtained in the absence of the vector potential: 
\bea
J_\mu(\xi)= \bar{\Psi}(\xi) \hat{Q}\gamma_\mu \Psi(\xi)=
\bar{\Psi}_{0}(\xi)e^{+i V \cdot \xi} \hat{Q} \gamma_\mu e^{-i V \cdot
  \xi} \Psi_{0}(\xi)=
\bar{\Psi}_{0}(\xi)\gamma_\mu\hat{Q}\Psi_{0}(\xi),\label{constr}
\eea
so that any exponential factors involving the vector potential are
cancelled. Thus the current commutator needed to compute the structure
function really only involves the fields $\Psi_0$ and
$\Psi_0^\dagger$, and the vector potential causes no explicit shift in
the argument.

We also note that \eq{reduce2} is consistent with both the baryon and
momentum sum rules. The latter follows from the feature that any
plus-momentum carried by the constant vector potential can be
associated with the plus-momentum of the nucleon. Thus effectively,
the constant vector potential carries no plus momentum
\cite{Miller:2001tg}.

Note that \eq{reduce2} results from using the same quark fields
throughout the calculation, so that mathematical consistency is
maintained. However, this consistency results from either allowing the
high-momentum struck quark to interact with the same vector potential
that influences quarks in the target ground state, or by ignoring the
vector potential altogether. Either procedure is questionable on
physical grounds.

\section{Evaluations and Physics }

Smith \& Miller \cite{millersmith} and Thomas {\it et al.}
\cite{Thomas:vt,Mineo:2003vc} each use an expression for the in-medium
quark distribution function \cite{explain}:
\bea \widetilde q(x)=
{1\over 2\sqrt{2}\pi}\int d\xi^-e^{-iP^+x\xi^-\over\sqrt{2}}
\langle P\mid \Psi^\dagger(\xi^-) P_+
\Psi(0)\vert P\rangle
\LARGE{\vert}_{\xi_\perp=0,\xi^+=0},\;
\label{reduce3}
\eea
where $P^+$ is the plus component of the momentum of the bound
nucleon. The phase factor $e^{-i V^+\xi^-}$ does not appear,
contradicting \eq{reduce1} \cite{oops}.  Thomas {\it et al.} then
exploit \eq{gauge} and use consistent normalization to obtain an
expression \bea q(x)={P^+\over P^+-V^+}q_0\left({P^+\over
    P^+-V^+}x-{V^+\over P^+}\right),\label{simple}\eea in which the
subscript 0 refers to the absence of the vector potential. The use of
\eq{simple} simplifies the evaluation of the distribution, but its use
is not necessary. One could alternatively evaluate \eq{reduce3}, and
this is the procedure of Smith \& Miller.

Both sets of authors avoid using the mathematically consistent, but
physically questionable \eq{reduce2}.  Indeed, the models of both sets
of authors are consistent with the baryon and momentum sum rules, and
produce results that are free of mathematical anomalies while
achieving reasonably good descriptions of a wide variety of phenomena.

That using \eq{reduce3} is a better procedure than using \eq{reduce2}
can be seen by examining \eq{green}. The presence of the phase factor
in \eq{reduce1} arises from allowing the struck quark to feel the same
mean-field vector-potential that the bound quarks experience. But such
mean fields are meant only to apply to the bound particles. Indeed,
the condition of asymptotic freedom mandates that the mean field
should not be felt by the struck quark for very large values of $Q^2$
\cite{quibble}.  Ignoring the phase factor of \eq{reduce1} is one way
to include the correct physics that the mean field should be
energy-dependent and should disappear at large momenta. The mean-field
should provide important effects for quarks in the target ground state
and should vanish for high-energy quarks.  Thus both sets of authors
\cite{Thomas:vt,Mineo:2003vc} and \cite{millersmith} use a physically
reasonable procedure. The rigorous task of deriving a vector potential
with the ability to account for both the high and low energy limits of
the quark self-energy in nuclear matter remains a task for the future.

\section*{Acknowledgement}
This work is supported in part by funds provided by the
U.S.~Department of Energy (D.O.E.)  under contract DE-FG-02-97ER41014.


\begin{thebibliography}{99}
\bibitem{leading} We do not discuss gluon radiation and the related
QCD evolution in the text. Both sets of authors interpret their models
as providing distributions at a low hadronic scale, and then use
evolution to reach the large values of $Q^2$.
\bibitem{oops} The words used in Ref.~\cite{millersmith} may have
  confused the reader, but \eq{reduce3} is used (for nucleons at rest
  in the medium).
\end{thebibliography}
\end{document}